# Soft Neutron Production - a window to the Final State Interactions in DIS

M. Strikman[a,b], M.G. Tverskoy[b], M.B.Zhalov[b]

[a] Pennsylvania State University, University Park, PA, USA
[d] S.Petersburg Nuclear Physics Institute, Russia

**Abstract:** We analyze the production of soft neutrons ($E_n \leq 10$ MeV) in deep inelastic scattering and conclude that the recent E-665 data indicate strong suppression of the final state interactions in DIS at high energies. Hence we suggest that studies of the energy dependence of the soft neutron yields would provide a sensitive probe of the dynamics of the final state interactions in DIS.

The final state interactions (FSI) in deep inelastic scattering allow us to probe the space-time evolution of strong interactions. Observing these effects is a challenging job, especially if one wants to study their energy dependence. The soft neutron yield ($E_n \leq 10 MeV$) is one of the very few observables which can be studied both at fixed target energies using standard detectors of low energy neutrons and at collider using a forward neutron calorimeter [1]

The mechanism of soft neutron production is reasonably well understood - such neutrons are produced by pre-equilibrium emission and evaporation from the residual nuclear system left after the cascade stage of the fast particle-nucleus interactions. In the case of heavy nuclei these neutrons provide the major channel of "cooling" of the residual system. For intermediate $A$, where the Coulomb barrier is not high, neutrons take away about half of the excitation energy, with the rest carried by the charged fragments.

The total neutron multiplicity is *approximately proportional* to the number of nucleons knocked out from the nucleus at the fast stage of interaction provided the number of the number of of knocked out nucleons is sufficiently small. Therefore, the measurement of such a multiplicity should provide *a global measure of the number of wounded nucleons*. The standard Monte-Carlo cascade codes describe the neutron yields in the intermediate energy reactions ($E_{inc} \leq 1 GeV$) reasonably well. We estimate the precision of current codes to be on the level $\sim 20\%$, which includes all possible reasonable variation of the input parameters such as the level density of nuclear states, etc.

Recently the E-665 collaboration [2] has reported the first data on the production of low energy neutrons with kinetic energies $E_n \leq 10$ MeV in DIS of high energy muons off a number of nuclear targets: D, C, Ca, Pb. Relatively small **average** multiplicity of such neutrons, $\langle N_n(A) \rangle$ was reported:

$$\langle N_n(Pb)(E_n \leq 10 \text{ MeV})\rangle \simeq 5 \pm 1 \pm 0.3. \qquad (1)$$

$$\langle N_n(Ca)(E_n \leq 10 \text{ MeV})\rangle \simeq 1.8 \pm 0.5 \pm 0.3. \qquad (2)$$

The second error reflects systematic errors associated with the method of background subtraction which assumed that background is the same as for scattering off the deuteron. Obviously this assumption is more important for the case of lighter targets where overall neutron multiplicity is smaller. Hence we will concentrate on the analysis of the Pb data and briefly comment on the Ca data.

---

[1] Note that for collider kinematics, soft neutrons have energies $\approx E_A/A$ and very small transverse momenta. As a result the H1 and ZEUS neutron detectors have nearly 100% acceptance for these neutrons. [1]

It appears that the only available high-energy data on production of soft neutrons were obtained at ITEP [4] for the incident protons with $1.4 \leq p_p^{inc} \leq 9$ GeV/c. We find that in the overlapping energy range: 7.5 MeV $\leq E_n \leq$ 10 MeV the spectra measured in the two experiments using lead target have similar shapes. However, the neutron multiplicity is much higher for the high-energy proton projectiles, which reflects the much larger number of wounded nucleons in $pA$ interactions. To compare the number of neutrons produced per wounded nucleon we calculated $\frac{1}{A\sigma_{tot}(aN)} \frac{d\sigma^3(a+A\to n+X)}{d^3p/E}$ for proton and muon projectiles, since in the Glauber type models without secondary interactions this quantity does not depend on the projectile [3]. We still found about a factor of two larger value of this quantity for the proton case, indicating that secondary interactions are more important for the proton projectile.

The DIS data were taken for small $x$ where some nuclear shadowing is observed. Shadowing leads to increase of the fraction of diffraction events in which no neutron is produced. However at the same time shadowing leads to increase of contribution of events where $\gamma^*$ inelastically interacts with $\geq$ two nucleons. It is easy to check that the overall effect is a small increase in the neutron multiplicity. So for the rough estimates we will neglect shadowing in the further analysis.

It is generally assumed that in DIS high-energy hadrons are formed beyond the nucleus and that only hadrons with energies $\leq$ few GeV are involved in FSI. The most conservative assumption seems to be that only recoiling nucleons reinteract in the nucleus. The spectrum of these nucleons can be approximated at small $x$ and not very large energies, where the triple Pomeron contribution is still small, as

$$\frac{1}{\sigma_{\gamma^*p}} \frac{zd\sigma^{\gamma^*+p\to N+X}}{dzd^2p_t} \propto \exp(Ap_t^2)\sqrt{z}, \tag{3}$$

with $\langle p_t^2 \rangle^{1/2} \sim 0.4$ GeV/c, see discussion in [3]. Here $z$ is Feynman $x$ for the nucleons. Therefore the average kinetic energies of the produced nucleons in the nucleus rest frame are of the order 300-400 MeV. Hardly any time formation arguments could be applicable in this case. So production of soft nucleons through the creation of the hole due to removal of one nucleon and subsequent interactions of this nucleon with the rest of the nucleus should be considered as a *lower limit* for the rate of the soft neutron production. This limit obviously does not include production of soft neutrons in the process of absorption of slow pions produced in the elementary lepton-nucleon DIS, for which time formation arguments do not apply as well.

To estimate this *lower limit* we used a Monte Carlo code for the intermediate energy hadron -nucleus interaction for all stages of the process: cascade, which includes processes of knock-out of nucleons, production and subsequent interaction of pions, pre-equilibrium emission and evaporation of neutrons and charged particles, see e.g. [5]. The total number of simulated events in every case was 50000. We focused on the case of scattering off Pb, since the muon data are more accurate in this case. We checked our version of the code using ITEP data [4]. and found a good agreement with these data, see Fig.1, confirming that we can trust predictions of the code with stated above accuracy of about 20 %. To calculate the rate of the soft neutron production we considered the following model: (i) a nucleon was removed from any point in the nucleus with a probability proportional to the nuclear density; (ii) An energy $W$ was assigned to it, (iii) Nucleon propagation in the nuclear medium and subsequent production of soft neutrons was modeled using the Monte-Carlo code tested using $pA$ data.

The multiplicity of the produced neutrons for different cutoffs in $E_n$ is shown in Fig.2. One can see that for kinetic energies of interest: 200 MeV $\leq W \leq$ 500 MeV the soft neutron multiplicities rather weakly depend on $W$. We estimate

$$\langle N_n(Pb)(E_n \leq 10 \text{ MeV})\rangle_{lowerlimit} = 6 \pm 1.5, \tag{4}$$

which is reasonably close to the experimental number of $5 \pm 1$. Qualitatively, the shape of the energy distribution is also reproduced. However, the shape for $E_n \geq 2 MeV$ is more sensitive to the energy spectrum of the knocked out nucleons. Thus a more detailed comparison with the data would require modeling of the initial spectrum of the knock out nucleons. This will be reported elsewhere. In the case of $\mu Ca$ scattering our limit for the same cuts is $N_n(Ca) \geq 1.2 \pm 0.2$ which is somewhat smaller than the number reported by E-665 (eq.2). However within the systematic errors the lower bound is not significantly below the data. Note also, that in the case of $A \ll 200$ one needs to perform further tests of the codes since deexcitation process is more complicated in this case: for Ca about 50% of the energy is released via emission of protons and heavier fragments, though for $A \geq 200$ neutrons carry practically away practically all excitation energy.

Let us briefly discuss how many more extra neutrons one could expect from the secondary interactions e.g. in the string models, where produced quark interacts with effective cross section $\sigma_{eff} \approx 20$ mb, for discussion of these models and references see [6]. The total number of nucleons knocked out by the quark can be easily calculated if the nucleon correlations in nuclei are neglected:

$$\Delta N(A) = \sigma_{eff} \frac{A-1}{2A^2} \int T^2(b)d^2b, \tag{5}$$

where $T(b) = \int_{-\infty}^{\infty} dz \rho_A(b,z)$, and nuclear density $\rho_A(\vec{r})$ is normalized to $\int d^3r \rho_A(\vec{r}) = A$. For $20 \leq A \leq 200$, $\Delta N(A) \propto A^{0.5}$. For $\sigma_{eff} = 20$ mb: $\Delta N(Ca) \simeq 0.7$, $\Delta N(Pb) \simeq 1.5$. Naively this would lead to increase of our lower bound estimate by a factor of 1.7 and 2.5 respectively. Clearly, this would be hard to accommodate at least in the case of scattering off Pb.

Thus we conclude that the E-665 data for neutron yield from Pb indicate substantial suppression of FSI. To reconcile this with the data at lower energies where substantial FSI effects were observed one would need to assume that FSI decrease with increase of the energy. This is consistent both with the trend of the E-665 data for $\langle N_n(Pb)(E_n \leq 10 \text{ MeV})\rangle$ to fall with increase of $E_{inc}$ and with decrease of the probability for the produced fast hadrons to reinteract with nuclei which was also observed by E665 [7].

We conclude that a systematic study of the energy dependence of the neutron yield starting from the incident energies available at HERMES is necessary. Since the absorption of the leading hadrons is expected to decrease rather rapidly with increase of $E_{inc}$ in the HERMES energy range, one can expect substantial decrease of the soft neutron yield in the HERMES kinematics. It would be interesting to study correlation between the soft neutron multiplicity and the spectrum of leading hadrons ($z$-distribution, $p_t$ broadening, etc). It is necessary also to repeat the E665 experiment at higher energies to check the rather amazing finding of this experiment of the low rate of production of soft neutrons. If the decrease with $E_{inc}$ and low neutron multiplicity at $E_\mu \geq 200$ GeV are confirmed, the soft neutrons would provide a perfect tool to look for relatively rare final state interactions in DIS at HERA collider at small $x$ by selecting events with much larger than average neutron multiplicity. The study of soft neutron production at the electron machines for $x \sim 1$ and $Q^2 = 0.5 - 1 GeV^2$ would be also of help to check dynamics of low-energy FSI.

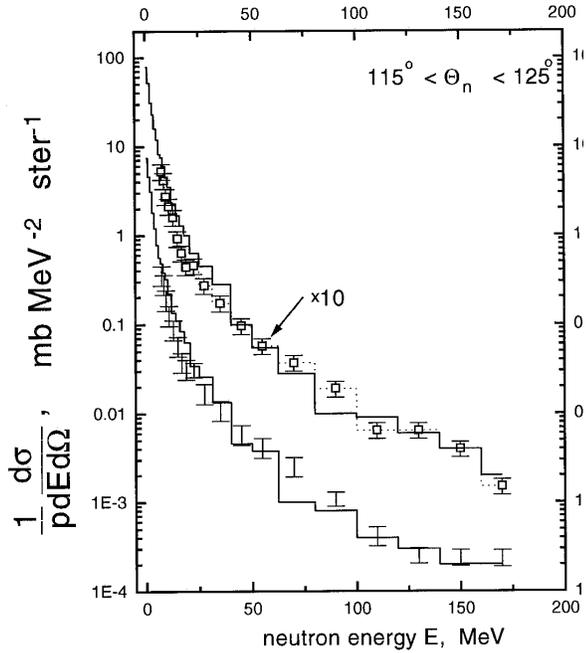 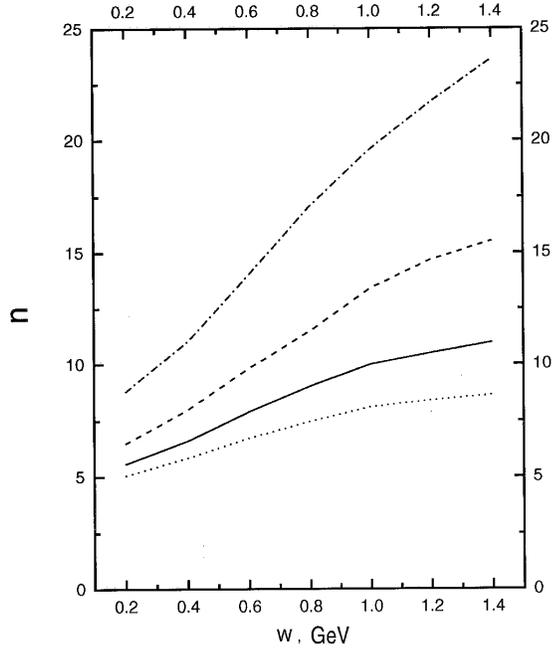

**Figure 1:** *Comparison of the results of the Monte Carlo cascade-evaporation calculation of the neutron spectra in $p + Pb \to n + X$ process -solid curves with the ITEP data [4] at $P_p = 1.4 GeV/c$ and $P_p = 2 GeV/c$ - dashed lines with open circles and squares.*

**Figure 2:** *Multiplicity of neutrons produced in the process where a nucleon with energy W was produced inside Pb for different energy intervals of the neutron energy. Dotted, solid, dashed curves are multiplicities of evaporated nucleons for $0 \leq T_n \leq 6$ MeV, $0 \leq T_n \leq 10$, $0 \leq T_n \leq 50$ MeV dashed-dotted curve is the total neutron multiplicity.*

We would like to thank K.Griffioen for discussions of the E-665 neutron data, G.van der Steenhoven for useful comments and B.Z.Kopeliovich for discussion of the string models. The research was supported by the DOE grant DE-FG02-93ER40771 and the ISF grant SAK 000.